# A separation of electrons and protons in the GAMMA-400 gamma-ray telescope

A.A. Leonov[b], A.M. Galper[a;b], V. Bonvicini[c], N.P. Topchiev[a], O. Adriani[d], R.L. Aptekar[e], I.V. Arkhangelskaja[b], A.I. Arkhangelskiy[b], L. Bergstrom[f], E. Berti[d], G. Bigongiari[g], S.G. Bobkov[h], M. Boezio[c], E.A. Bogomolov[e], S. Bonechi[g], M. Bongi[d], S. Bottai[d], G. Castellini[j], P.W. Cattaneo[k], P. Cumani[c], G.L. Dedenko[b], C. De Donato[l], V.A. Dogiel[a], M.S. Gorbunov[h], Yu.V. Gusakov[a], B.I. Hnatyk[n], V.V. Kadilin[b], V.A. Kaplin[b], A.A. Kaplun[b], M.D. Kheymits[b], V.E. Korepanov[o], J. Larsson[m], V.A. Loginov[b], F. Longo[c], P. Maestro[g], P.S. Marrocchesi[g], V.V. Mikhailov[b], E. Mocchiutti[c], A.A. Moiseev[p], N. Mori[d], I.V. Moskalenko[q], P.Yu. Naumov[b], P. Papini[d], M. Pearce[m], P. Picozza[l], A.V. Popov[h], A. Rappoldi[k], S. Ricciarini[j], M.F. Runtso[b], F. Ryde[m], O.V. Serdin[h], R. Sparvoli[l], P. Spillantini[d], S.I. Suchkov[a], M. Tavani[r], A.A. Taraskin[b], A. Tiberio[d], E.M. Tyurin[b], M.V. Ulanov[e], A. Vacchi[c], E. Vannuccini[d], G.I. Vasilyev[e], Yu.T. Yurkin[b], N. Zampa[c], V.N. Zirakashvili[s] and V.G. Zverev[b]

[a] Lebedev Physical Institute, Russian Academy of Sciences, Moscow, Russia
[b] National Research Nuclear University MEPhI, Moscow, Russia
[c] INFN, Sezione di Trieste and Physics Department of University of Trieste, Trieste, Italy
[d] INFN, Sezione di Firenze and Physics Department of University of Florence, Firenze, Italy
[e] Ioffe Institute, Russian Academy of Sciences, St. Petersburg, Russia
[f] Stockholm University, Department of Physics; and the Oskar Klein Centre, AlbaNova University Center, Stockholm, Sweden
[g] Department of Physical Sciences, Earth and Environment, University of Siena and INFN, Sezione di Pisa, Italy
[h] Scientific Research Institute for System Analysis, Russian Academy of Sciences, Moscow, Russia
[i] Research Institute for Electromechanics, Istra, Moscow region, Russia
[j] IFAC- CNR and Istituto Nazionale di Fisica Nucleare, Sezione di Firenze, Firenze, Italy
[k] INFN, Sezione di Pavia, Pavia, Italy
[l] INFN, Sezione di Roma 2 and Physics Department of University of Rome Tor Vergata, Italy
[m] KTH Royal Institute of Technology, Department of Physics; and the Oskar Klein Centre, AlbaNova University Center, Stockholm, Sweden
[n] Taras Shevchenko National University of Kyiv, Kyiv, Ukraine
[o] Lviv Center of Institute of Space Research, Lviv, Ukraine
[p] CRESST/GSFC and University of Maryland, College Park, Maryland, USA
[q] Hansen Experimental Physics Laboratory and Kavli Institute for Particle Astrophysics and Cosmology, Stanford University, Stanford, USA
[r] Istituto Nazionale di Astrofisica IASF and Physics Department of University of Rome Tor Vergata, Rome, Italy
[s] Pushkov Institute of Terrestrial Magnetism, Ionosphere, and Radiowave Propagation, Troitsk, Moscow region, Russia

**Abstract**

The GAMMA-400 gamma-ray telescope is intended to measure the fluxes of gamma rays and cosmic-ray electrons and positrons in the energy range from 100 MeV to several TeV. Such measurements concern with the following scientific goals: search for signatures of dark matter, investigation of gamma-ray point and extended sources, studies of the energy spectra of Galactic and extragalactic diffuse emission, studies of gamma-ray bursts and gamma-ray emission from the active Sun, as well as high-precision measurements of spectra of high-energy electrons and positrons, protons, and nuclei up to the knee.

The main components of cosmic rays are protons and helium nuclei, whereas the part of lepton component in the total flux is $\sim 10^{-3}$ for high energies. In present paper, the capability of the GAMMA-400 gamma-ray telescope to distinguish electrons and positrons from protons in cosmic rays is investigated. The individual contribution to the proton rejection is studied for each detector system of the GAMMA-400 gamma-ray telescope. Using combined information from all detector systems allow us to provide the proton rejection from electrons with a factor of $\sim 4 \times 10^5$ for vertical incident particles and $\sim 3 \times 10^5$ for particles with initial inclination of 30 degrees. The calculations were performed for the electron energy range from 50 GeV to 1 TeV.

**1. Introduction**

The GAMMA-400 instrument has developed to address a broad range of scientific goals, such as search for signatures of dark matter, studies of Galactic and extragalactic gamma-ray sources, Galactic and extragalactic diffuse emission, gamma-ray bursts, as well as high-precision measurements of spectra of cosmic-ray electrons, positrons, and nuclei (Galper et al., 2013). In this paper, the electron and proton separation methods for the GAMMA-400 gamma-ray telescope are presented.

The detector systems used in the satellite-borne experiments and devoted to study high-energy cosmic rays have a possibility not only to measure energies, but also to identify protons and electrons. This identification is usually based on a comparison of longitudinal and transversal shower profiles and the total energy deposition in a calorimeter system taking into account that electromagnetic and hadronic showers have different spatial and energy topology form (Fabjan and Gianotti, 2003). Moreover, the number of neutrons generated in the electromagnetic cascade is much less than that in the hadronic cascade. The neutron detection essentially improves separation of electrons from protons (Adriani et al., 2009; Stozhkov et al., 2005).

The combined information from all detector systems provides the rejection factor for vertical and inclined protons better than $10^5$.

## 2. The detector systems of the GAMMA-400 gamma-ray telescope

The GAMMA-400 physical scheme is shown in Fig. 1. GAMMA-400 consists of scintillation anticoincidence top and lateral detectors (AC top and AC lat), converter-tracker (C) with 8 layers of double (x, y) silicon strip coordinate detectors (pitch of 0.1 mm) interleaved with tungsten conversion foils and 2 layers of double (x, y) silicon strip coordinate detectors without tungsten at the bottom, scintillation detectors (S1 and S2) of time-of-flight system (ToF), calorimeter from two parts (CC1 and CC2), lateral detectors (LD), scintillation detectors (S3 and S4) and neutron detector (ND) to separate hadronic and electromagnetic showers.

The anticoincidence detectors surrounding the converter-tracker are served to distinguish gamma-ray events from the much more numerous charged-particle events. Converter-tracker information is applied to precisely determine the direction of each incident particle and calorimeter measurements are used to determine its energy. All scintillation detectors consist from two independent layers. Each layer has thickness of 1 cm. The time-of-flight system, where detectors S1 and S2 are separated by 500 mm, determines the top-down direction of arriving particle. Additional scintillation detectors S3 and S4 improve hadronic and electromagnetic showers separation.

The imaging calorimeter CC1 consists of 2 layers of double (x, y) silicon strip coordinate detectors (pitch of 0.1 mm) interleaved with planes from CsI(Tl) crystals, and the electromagnetic calorimeter CC2 consists of CsI(Tl) crystals with the dimensions of 36 mm × 36 mm × 430 mm. The total converter-tracker thickness is about 1 $X_0$ ($X_0$ is the radiation length). The thicknesses of CC1 and CC2 are 3 $X_0$ and 22 $X_0$, respectively. The total calorimeter thickness is 25 $X_0$ or 1.2 $\lambda_0$ ($\lambda_0$ is nuclear interaction length). Using thick calorimeter allows us to extend the energy range up to several TeV and to reach the energy resolution better than 1% above 100 GeV.

The presence of the signal from AC detectors and the absence of the signal from AC detectors are necessary to detect charged particles and gamma rays, respectively.

## 3. Methods to reject protons from electrons using the GAMMA-400 gamma-ray telescope systems

The rejection of protons from electrons using the separation methods was performed with the GEANT4 simulation toolkit software (http://geant4.cern.ch). Protons produce the main background, when detecting cosmic-ray electrons, since a fraction of the lepton component is ~$10^{-3}$ of the total cosmic-ray flux for high energies. The main trigger of the gamma-ray telescope includes the signals from time-of-flight scintillation detectors S1 and S2, when the signal in S1 has to be generated before the signal in S2. To reject protons from electrons the GAMMA-400 instrument information from ND, S4, S3, S2, CC1, and CC2 is used.

Proton rejection factor should take into account the final proton contamination in the resulting reconstructed electron flux. It could be calculated by applying the electron selections separately to the proton and the electron spectra and by constructing energy histogram with surviving electrons and protons in each energy bin. This will immediately provide the electron detection efficiency, which can be transferred in the effective geometric factor for electrons, and proton contamination. After that, in order to calculate the flux, it is necessary just simply to multiply the total flux (electrons + residual protons) by the simulated value.

The rejection factor to separate protons from electrons with energy 100 GeV is calculated as the ratio of the number of initial protons with energy more than 100 GeV (the proton energy spectral index is equal to -2.7), assuming that the proton energy spectrum power is -2.7, to the number of events identified as electrons with energy 100±2 GeV (taking into account that the GAMMA-400 energy resolution is equal to about 2%).

Similar analyses of electron measurements and proton rejection techniques using calorimeters that nearly fully contain electromagnetic showers, but are thin in hadronic interaction lengths, as the GAMMA-400 calorimeter is, have been carried out many times. These have consistently indicated that the most powerful methods to reject proton events are the initial point of the shower, the lateral distribution of particles in the calorimeter, and the longitudinal shower development.

Taking into account the GAMMA-400 instrument structure, the information from CC2 can be applied for rejection using lateral distribution of particles in the calorimeter; S4 responses allow us to reduce proton contamination from the longitudinal shower development. The initial point of the shower is utilized from the information of S2, S3, and CC1 detectors. The information from neutron detector also essentially improves separation of electrons from protons, especially for the energy more, than 50 GeV (Stozhkov et al., 2005).

Firstly, the contamination for vertical incident protons is evaluated. All processed criteria to suppress protons are based on selecting cutoffs to distinguish proton and electron events. The location of the cutoff for each criterion is selected in order to retain 98% of electrons. Totally 25 cutoffs are used to reject protons. Taking into account presented selection ~30% of electrons are also lost due to proton rejection. The description of the criteria is ordered according to their own rejection power, excepting the number of neutrons in ND. Because the ND efficiency in view of neutrons detection will be the purpose of the next more detailed calculations, the power of the ND criterion can be considered as upper estimation of own rejection. This is not critical due to the fact that the refusal of this criterion does not reduce the total rejection more than 2 times.

The information from S4 provides the strongest own rejection factor for protons. This rejection concerns with the difference in the attenuation for hadronic and electromagnetic cascades. Electromagnetic shower initiated by the electron with initial energy of ~ 100 GeV is fully contained inside the calorimeter with the thickness 25 $X_0$ or 1.2 $\lambda_0$. Such criterion was used in the PAMELA experiment data analysis (Karelin et al., 2013).

For correct calculations of own rejection the width of distribution of signal amplitudes from electrons and protons is the mainly important parameter. In the GEANT 4 simulations, the value of scintillation detector response is formed only by ionization losses of the particle inside the detector. To take into account the efficiency of releasing energy conversion into the light flash, the efficiency of light collection, and the efficiency of the light to electric signal transformation in a photomultiplier, the data of beam test calibration were used (Boezio et al., 2004). In Fig. 2, the solid line presents the dependence of ratio of RMS to the mean signal in scintillation detector from the energy of electrons in the CERN calibration test beam. This calibration was performed in the frame of PAMELA experiment for the scintillation detector similar to the one intended to apply in the GAMMA-400 instrument. The GEANT4 simulation results are also shown in Fig. 2 by dashed line. Comparing the results of calibration and simulation, it is possible to introduce additional spreading parameter for the simulation data. The value of a signal in the scintillation detector is calculated as:
$E=E_0+E'$,
where $E_0$ is the energy release due to ionization losses,
$E'$ is the spreading signal calculated from the Gauss distribution with
$\sigma \approx 0.33 \times E_0^{-4} \times (10^{-4}) \times E_0^2$,
$E_0$ is determined in MIPs (MIP is minimum ionizing particle, 1 MIP is equal to ~ 2 MeV for S4).

Such spreading of the simulated response was performed for all scintillation detectors: S2, S3, and S4.

The distributions for signals in S4 from initial electrons and protons are shown in Figs. 3a and 3c. Selecting events with signals in S4 less than 2.7 MIP, it is possible to suppress protons with a factor of 100.

Additional rejection is obtained when analyzing total CC2 signals. The CC2 calorimeter contains CsI(Tl) square crystals with cross dimension of 36 mm and longitudinal dimension of 430 mm. The criterion is based on the difference of the transversal size for hadronic and electromagnetic showers. Such topology difference was successfully applied with the calorimeter data in the PAMELA mission to separate electrons from antiproton sample and positrons from proton sample (Karelin et al., 2013; Menn et al., 2013). The distributions for the ratio of a signal in the crystal containing cascade axis to the value of total signal in CC2 for initial electrons and protons are compared (Figs 3b and 3d). Using the distribution for the initial electrons, the values of two cutoffs are determined as 71.3% and 74.4%. For the proton distribution, only events placed between these electron cutoffs are retained. Applying this rejection provides the rejection factor of ~30.

The differences in proton and electron cascade transversal size are also used when analyzing information from silicon strips in CC1. The distributions for RMS of coordinates of strips with signals for initial electrons and protons are shown in Figs. 4a and 4b. For the proton distribution, only events placed between the cutoffs of 0.3 cm and 8 cm obtained from the electron distribution are retained. Applying this rejection provides the rejection factor of ~6.

To take into account the fact that the hadronic cascade begins to develop deeper inside the instrument, than the electromagnetic shower, the signals in detector systems CC1 (crystal), S2 and S3 are considered, using the fact that the thickness of material just above these detectors is less than 4 $X_0$. The distributions of signals in second layer of crystal CsI(Tl) from CC1 and of signals in S3 from initial electrons and protons are shown in Fig. 5. For proton-induced cascades, there are a lot of events with small signal amplitude. To reject such events the cutoff from electron induced cascades distribution is determined. The events from proton induced distribution with the value of signal less than this cutoff are rejected. This criterion allows us to suppress protons with the factor of 3 for each CsI(Tl) crystal from CC1 and with the factor of 2 for each scintillation detector S2 and S3.

The ND contribution in the rejection factor for protons in the GAMMA-400 telescope is considered with significantly different number of neutrons generated in the electromagnetic and hadronic cascades. In cascades, induced by protons, the generation of neutrons is more intensive than in the electromagnetic shower. The source of neutrons in cascades, induced by electrons, concerns mainly with generation of gamma rays in those cascades with energy close to 17 MeV. These gamma rays, in turn, could generate neutrons in the Giant resonance reaction. Analyzing information from the neutron detector placed just under the CC2 calorimeter, it is possible to suppress protons by the factor of 400. The distributions for number of neutrons at the entrance of ND from initial electrons and protons are shown in Fig. 6. The cutoff for the number of neutrons to separate protons is equal to 60. The efficiency of neutron detection is not taken into account in the present simulation, but will be the purpose of the next more detailed calculations. The power of this criterion can be considered as upper estimation of own rejection.

All above discussed proton rejection criteria were considered separately from each other. Using all criteria in the combination, it is possible to obtain rejection factor for protons equal to $(4\pm0.4)\times10^5$. Table 1 contains the information about own rejection factor of each criterion (without other) and the values of the total rejection factor decreasing in the case of the refusal from given criterion. All of presented criteria are strongly dependent that is confirmed by the values in the right column. The refusal from any given criterion, especially the strongest from S4, CC2, and CC1 detectors, reduces the total rejection factor significantly less than dividing with own rejection number of this criterion.

The same proton rejection algorithm is applied for inclined incident protons. The values of cutoffs for rejection procedure are changed. The results of calculation for $30^0$ incident angles of protons give the following rejection cutoffs. To suppress inclined protons only events with the number of neutrons in ND less than 75 are retained. To reject inclined protons only events with signals in S4 less than 1.1 MIP are considered. In CC2, the cascade axis from initial inclined particle crosses mainly three columns of CsI(Tl) crystals. The location of the axis of the cascade in CC2 is presented in Fig. 7. The transversal coordinates of lateral columns from the CsI(Tl) crystal, the column height and transversal coordinates of the cascade axis are also shown. For cascades induced by inclined electrons, the column number 4 (Fig. 7) with the maximum energy release contains only 30% of total energy deposited in CC2 (Fig. 8) instead of 70% for vertical particles (Figs. 3b and 3d). Using only distributions for the ratio of the signal in the crystal with maximum energy release to the value of total signal in CC2, it is possible to obtain the rejection factor for inclined protons not more than 14, instead of 30 for vertical particles. The corresponding cutoffs are 22.2% and 35.0%. But taking into account the distributions in the two other columns (Figs. 9a, 9c with cutoffs of 7.7% and 25.8%, Figs. 9b, 9d with cutoffs 13.5% and 27.2%), the value of the rejection factor for protons can be increased to 28.

From the distribution for RMS of coordinates of strips with signals in CC1 for initial inclined electrons the cutoff values of 0.7 cm and 7.4 cm are obtained. A new rejection cutoffs for inclined particles for each layer of the CsI(Tl) crystal from CC1, for scintillation detectors S2 and S3 will be 16.4 MIP, 53.3 MIP, 3.5 MIP, and 65.2 MIP, respectively. Using all criteria in the combination, it is possible to obtain the rejection factor for protons with initial incident angle of $30^0$ equal to $(3\pm0.4)\times10^5$.

The same calculations were performed for the energy range from 50 GeV to 1 TeV. Table 2 contains the information concerning the rejection factor to separate protons from electrons in this energy range.

The absolute values of all described cutoffs are increased with the gamma-ray initial energy. As an example, the dependence of the cutoffs in upper and bottom layers of the S4 detector versus the gamma-ray initial energy is presented in Fig. 10. The value of the cutoff in MIPs for the each energy is selected in order to retain 98% of electrons (Figs. 3a and 3c).

## 4. Conclusion

Using the combined information from all detector systems of the GAMMA-400 gamma-ray telescope, it is possible to provide effective rejection of protons from electrons. The proposed methods are based on the difference of the development of hadronic and electromagnetic showers inside the instrument. It was shown that the rejection factor for vertical and inclined protons is several times better than $10^5$. Such kind of separation extends in the energy range from 50 GeV to 1 TeV.

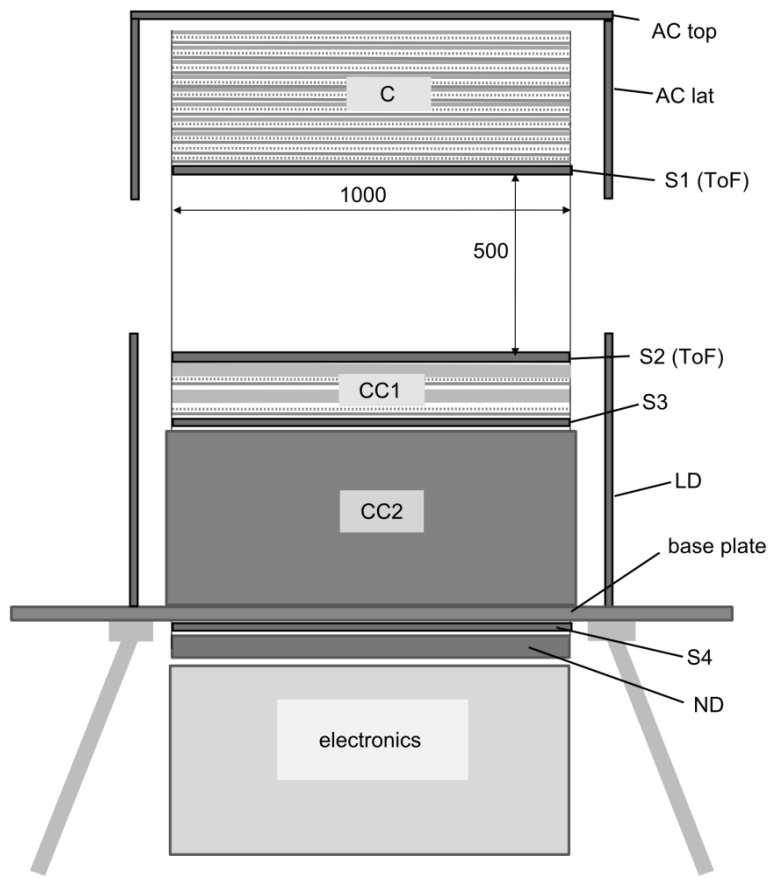

*Figure. 1. The GAMMA-400 physical scheme.*
*AC top is the top anticoincidence detector, AC lat are lateral anticoincidence detectors, C is the converter-tracker; S1 (ToF) and S2 (ToF) are scintillation detectors of the time-of-flight system, CC1 and CC2 are coordinate-sensitive and electromagnetic calorimeter, S3 and S4 are scintillation detectors, ND is the neutron detector.*

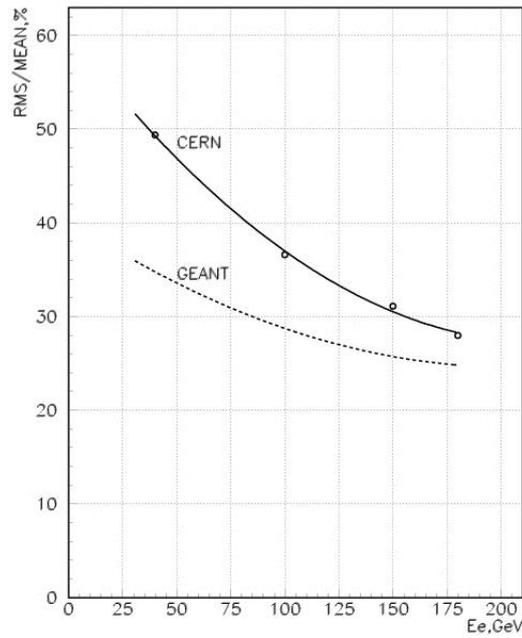

*Figure 2. The dependence of ratio of RMS to the average signal in scintillation detector from the energy of electrons in the CERN calibration test beam (solid line). The simulated dependence of ratio of RMS to the average signal in scintillation detector from the energy of electrons (dashed line).*

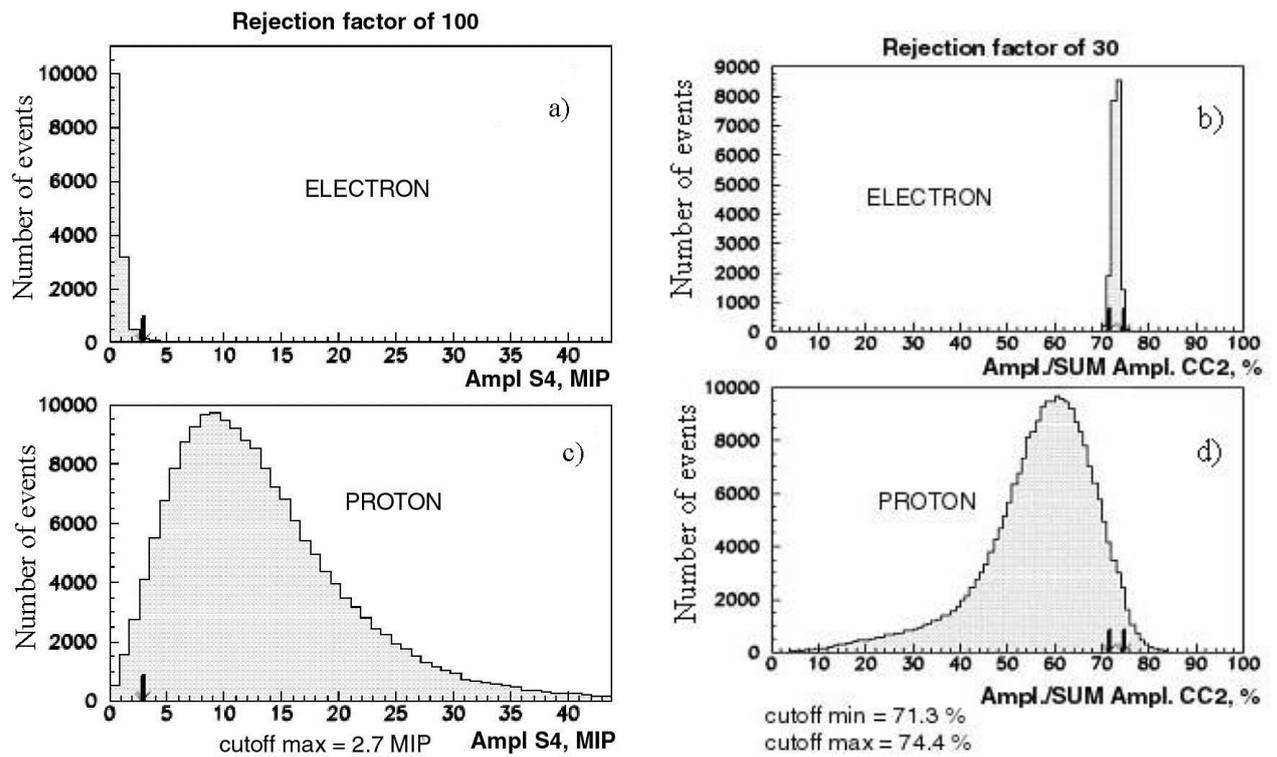

*Figure 3. The distributions for signals in S4 from initial electrons (a) and protons (c). The distributions for the ratio of a signal in the crystal containing the cascade axis to the value of total signal in CC2 for initial electrons (b) and protons (d).*

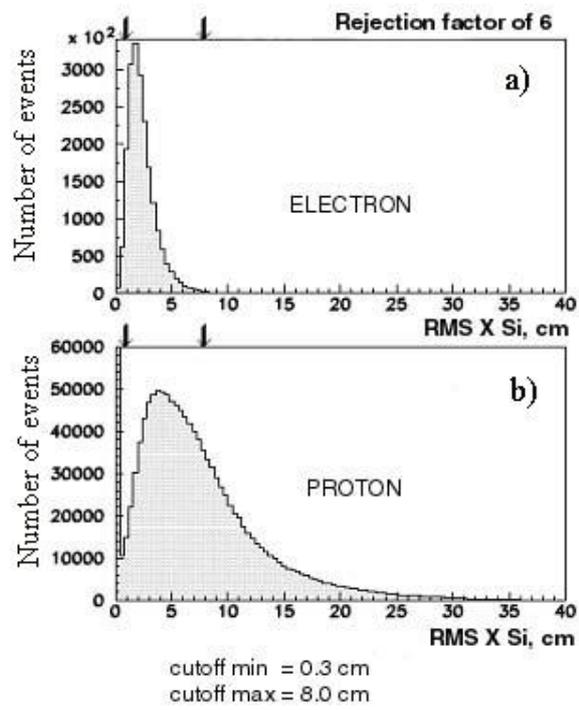

*Figure 4. The distributions for RMS of coordinates of strips with signals in CC1 for initial electrons (a) and protons (b).*

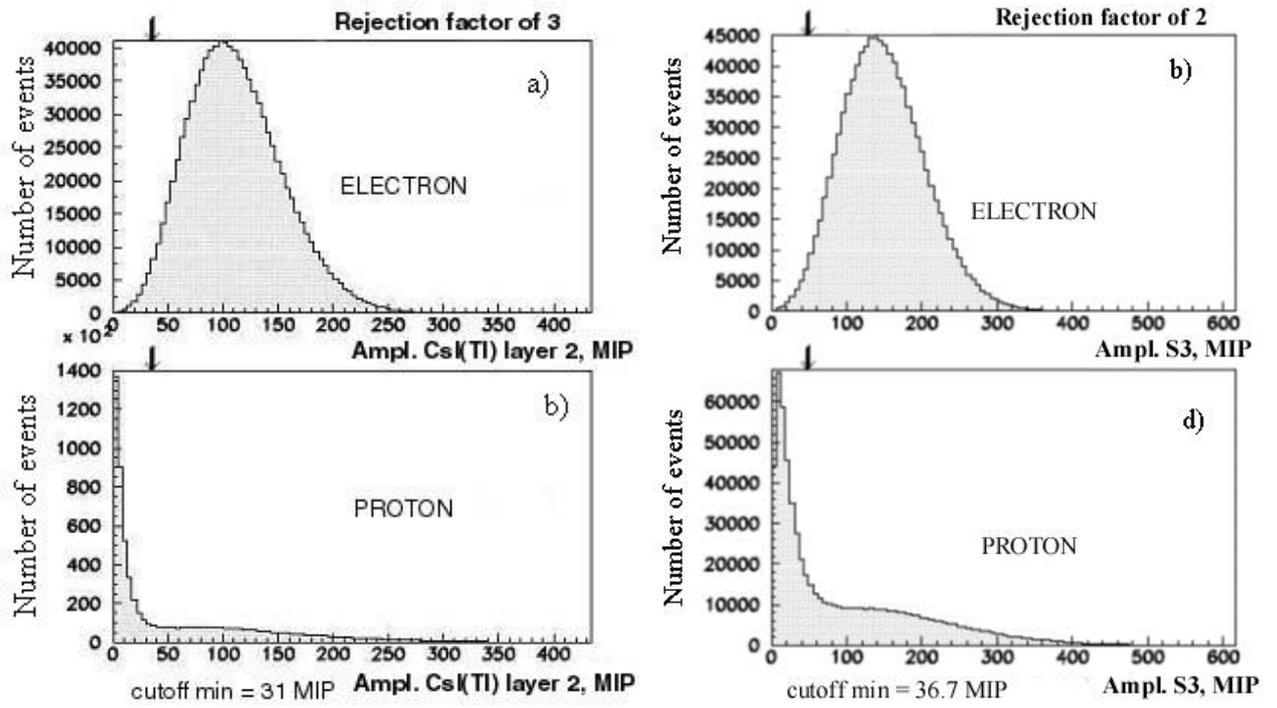

*Figure 5. The distributions of signals in second crystal CsI(Tl) of CC1 from initial electrons (a) and protons (c). The distributions of signals in S3 from initial electrons (b) and protons (d).*

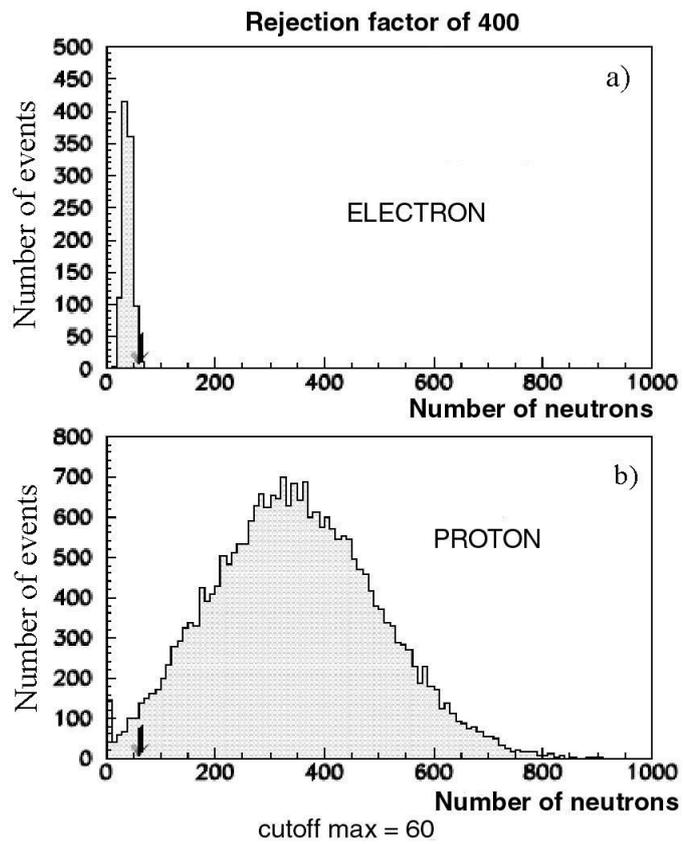

*Figure 6. The distributions for number of neutrons at the entrance of ND from initial electrons (a) and protons (b).*

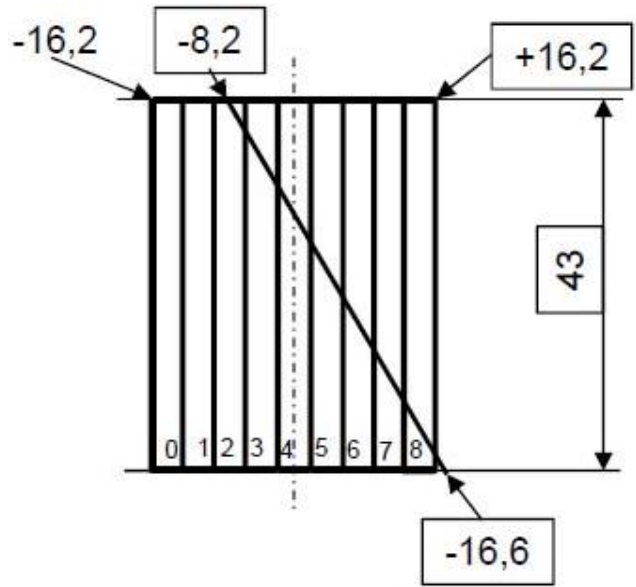

*Figure 7. The location of the axis of the proton induced cascade in CC2 (sloping line). The transversal coordinates of lateral columns from CsI(Tl) crystal, the column height and transversal coordinates of the cascade axis are shown.*

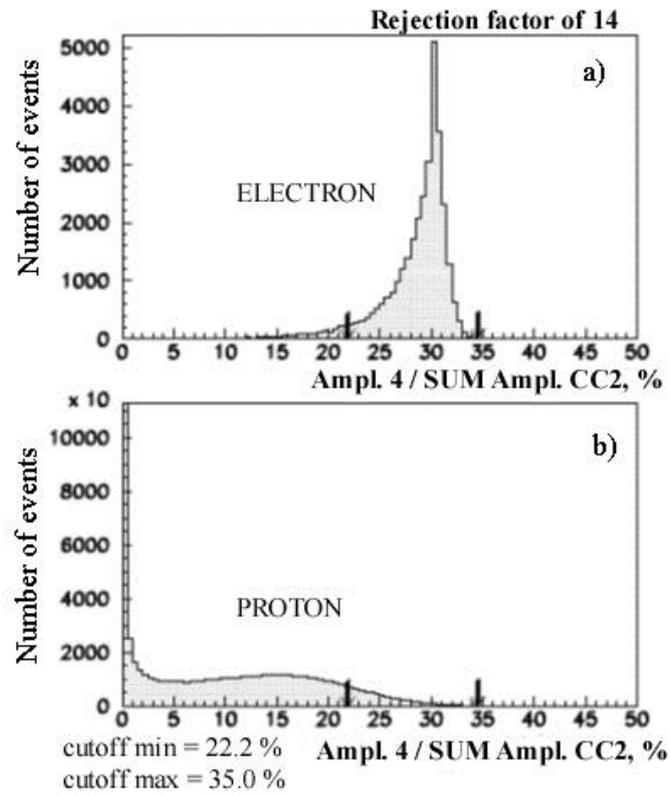

*Figure 8. The distributions for the ratio of signal in the crystal containing the cascade axis with maximum energy release to the value of total signal in CC2 for inclined incident electrons (a) and inclined incident protons (b).*

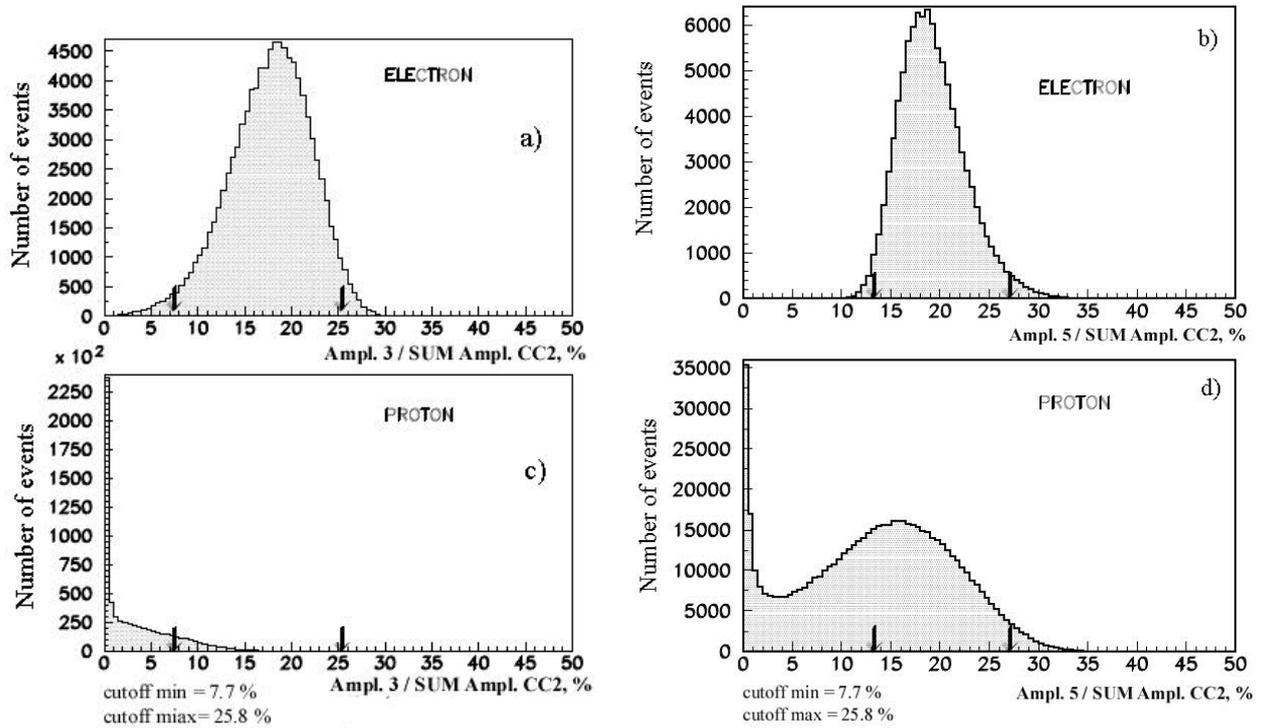

*Figure 9. The distributions for the ratio of signal in the crystal no. 3 from Fig. 6 to the value of total signal in CC2 for initial top-down electrons (a) and top-down protons (c). The distributions for the ratio of signal in the crystal no. 5 from Fig. 6 to the value of total signal in CC2 for initial top-down electrons (b) and top-down protons (d).*

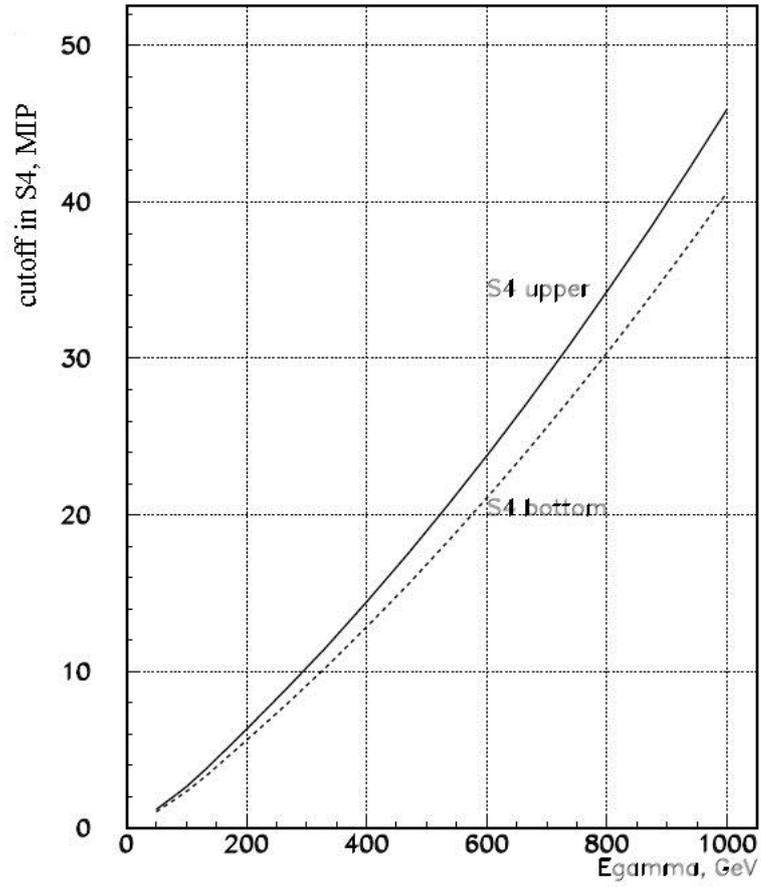

*Figure 10. The dependence of the cutoffs in upper (solid line) and bottom (dotted line) layers of the S4 detector versus the gamma-ray initial energy.*

*Table 1. Own rejection factor for protons of each detector system (without other) and the values of total rejection factor decreasing in the case of the refusal of given rejection.*

| Detector system, number of cutoffs | Own rejection factor | Total rejection factor decreasing |
|---|---|---|
| S4 (2 cutoffs: 1 cutoff for each scintillation layer) | 100 | 1.7 |
| CC2 (2 cutoffs) | 30 | 2.6 |
| Strips in CC1 (4 cutoffs: 2 cutoffs for each X or Y silicon strip) | 6 | 1.2 |
| CsI(Tl) from CC1 (2 cutoffs: 1 cutoff for each layer of CsI(Tl) crystal) | 3 | 1.3 |
| S2, S3 (4 cutoffs: 2 cutoffs for each detector) | 2 | 1.3 |
| ND (1 cutoff) | 400 (upper limit) | 2 |

*Table 2. Total rejection factor to separate protons from electrons in energy range from 50 GeV to 1 TeV.*

| Energy, GeV | Total rejection factor |
|---|---|
| 50 | $(12.8 \pm 2) \times 10^5$ |
| 100 | $(4.0 \pm 0.4) \times 10^5$ |
| 200 | $(5.0 \pm 0.7) \times 10^5$ |
| 1000 | $(4.1 \pm 0.7) \times 10^5$ |